\documentclass[10pt]{iopart}
\usepackage{iopams}
\usepackage{graphicx}
\usepackage{bm}
\usepackage[utf8]{inputenc}
\usepackage[T1]{fontenc}
\usepackage{comment}
\usepackage{siunitx}
\usepackage{cite}
\usepackage{color}
\usepackage{cancel}

\newcommand{\vect}[1]{\ensuremath{\mathbf #1}}
\newcommand{\hatt}[1]{\ensuremath\hat{{\mathbf #1}}}

\begin{document}
\title{Generation of super-stable vector modes using on-axis complex-amplitude modulation}


\author{Valeria Rodr\'iguez-Fajardo$^1$, Fernanda Arvizu$^2$, Dayver Daza-Salgado$^2$, Benjamin Perez-Garcia$^3$, Carmelo Rosales-Guzm\'an$^2$}
\address{$^1$Department of Physics and Astronomy, Colgate University, 13 Oak Drive, Hamilton, NY 13346, United States of America}
\address{$^2$Centro de Investigaciones en Óptica, A.C., Loma del Bosque 115, Colonia Lomas del Campestre, 37150 León, Gto., Mexico}
\address{$^3$Photonics and Mathematical Optics Group, Tecnologico de Monterrey, Monterrey 64849, Mexico.}

\ead{vrodriguezfa@gmail.com}

\vspace{10pt}
\begin{indented}
\item[]\today
\end{indented}

%
%
%

\begin{abstract}
In this manuscript, we propose the generation of complex vector beams with high quality and stability based on a novel approach that relies on the combination of two techniques that seem incompatible at first glance. The first is Complex Amplitude Modulation (CAM), which produces scalar structured light fields in phase and amplitude with high accuracy. The second is on-axis modulation for the generation of vector beams, a method that requires phase-only holograms, therefore yielding beams of reduced quality. More precisely, the idea behind our technique is to send the shaped light produced by CAM co-axially to the zeroth order, rather than to the first order, as commonly done. We describe our technique, explaining the generation of the hologram and experimental setup to isolate the desired vector mode, and then present experimental results that corroborate our approach. We first address the quality of the generated beams using Stokes polarimetry to reconstruct their transverse polarisation distribution, and then compare their stability against the same mode produced using a popular interferometric method. Our vector beams are of good quality and remarkably stable, two qualities that we expect will appeal to the community working with vector modes.
\end{abstract}

%
%
%
%
\ioptwocol

\section{Introduction}
Complex vector beams, non-separable in their spatial and polarisation degrees of freedom, are of great relevance at both the fundamental and applications aspects\cite{Rosales2018Review,Roadmap,Zhan2009,forbes_structured_2021}. From a fundamental point of view, vector beams have gained popularity as classical analogous of quantum-entangled states, due to the mathematical similarity in which both are expressed \cite{Shen2022,toninelli2019concepts}. From the applied side, vector beams are paving the way for applications in fields as diverse as optical communications, optical tweezers, and optical metrology, to name a few \cite{Ndagano2018,Yuanjietweezers2021,Ndagano2017,BergJohansen2015,hu2019,Fang2021}.

In this context, the generation of arbitrary vector beams with high purity and stability in their polarisation distribution is highly desired. Several methods have been proposed for their generation, amongst these, the use of digital devices has gained popularity since they are more flexible and versatile in comparison to other devices \cite{Hu2022,Scholes2019,Rong2014,Moreno2015,Wang2007,Rosales2017,Gong2014,Perez-Garcia2017,Hu2021Random,Maurer2007}. Techniques employing them commonly involve the implementation of all sort of interferometric setups, allowing to manipulate independently both the polarisation and spatial degrees of freedom of light \cite{Rosales2020,Mitchell2016,Ren2015}. High purity can be achieved by simultaneously shaping the phase and amplitude of a given light field. A popular technique, which can be implemented using phase-only Spatial Light Modulators (SLMs), is known as Complex Amplitude Modulation (CAM), which produces some of the most accurate results \cite{arrizon2007,Clark2016,Rosales2017}. The main idea behind CAM is the codification of both the amplitude and phase information of the desired light field into a digital hologram that requires a sinusoidal grating to separate the desired modulated light, commonly lying in the first diffraction order, from the undesired one. In this case, the zeroth diffraction order is always the brightest, in contrast to phase-only modulation, where the first order can be the most bright. To achieve high stability in the generation of vector beams, a possible alternative is to avoid interferometric arrays and to use on-axis modulation. Pioneering approaches have already implemented these methods, by taking advantage of the fact that SLMs can only modulate one linear polarisation, commonly horizontal. In essence, a diagonally polarized beam impinging onto an SLM gets shaped only along its horizontal polarisation component, leaving the vertical unmodified. Both polarisation components are then rotated to the corresponding orthogonal polarisation by means of a half-wave plate and sent to another SLM, where the previously un-modulated light gets now modulated and the light modulated by the first SLM is left intact, thereby producing a vector beam. Crucially, this approach requires the beam to always propagate on-axis, therefore a diffraction grating is avoided, and the use of phase-only holograms to produce on-axis spatial modulation is required, which produces beams of subpar quality. 

It is thus clear that in order to generate high-quality vector beams with high stability, we should combine CAM with on-axis generation, two techniques that seem incompatible at first glance. It is the main purpose of this manuscript to demonstrate that these two techniques can indeed be combined to generate vector beams with high quality and stability. We first explain the idea behind our technique and demonstrate it experimentally afterward. We assess the quality of the generated vector beams, by reconstructing their transverse polarisation distribution through Stokes polarimetry, and stability through studying the variations of intensity over time.

\section{The principle} 
\begin{figure*}[tbh]
    \centering
    \includegraphics[width=172mm]{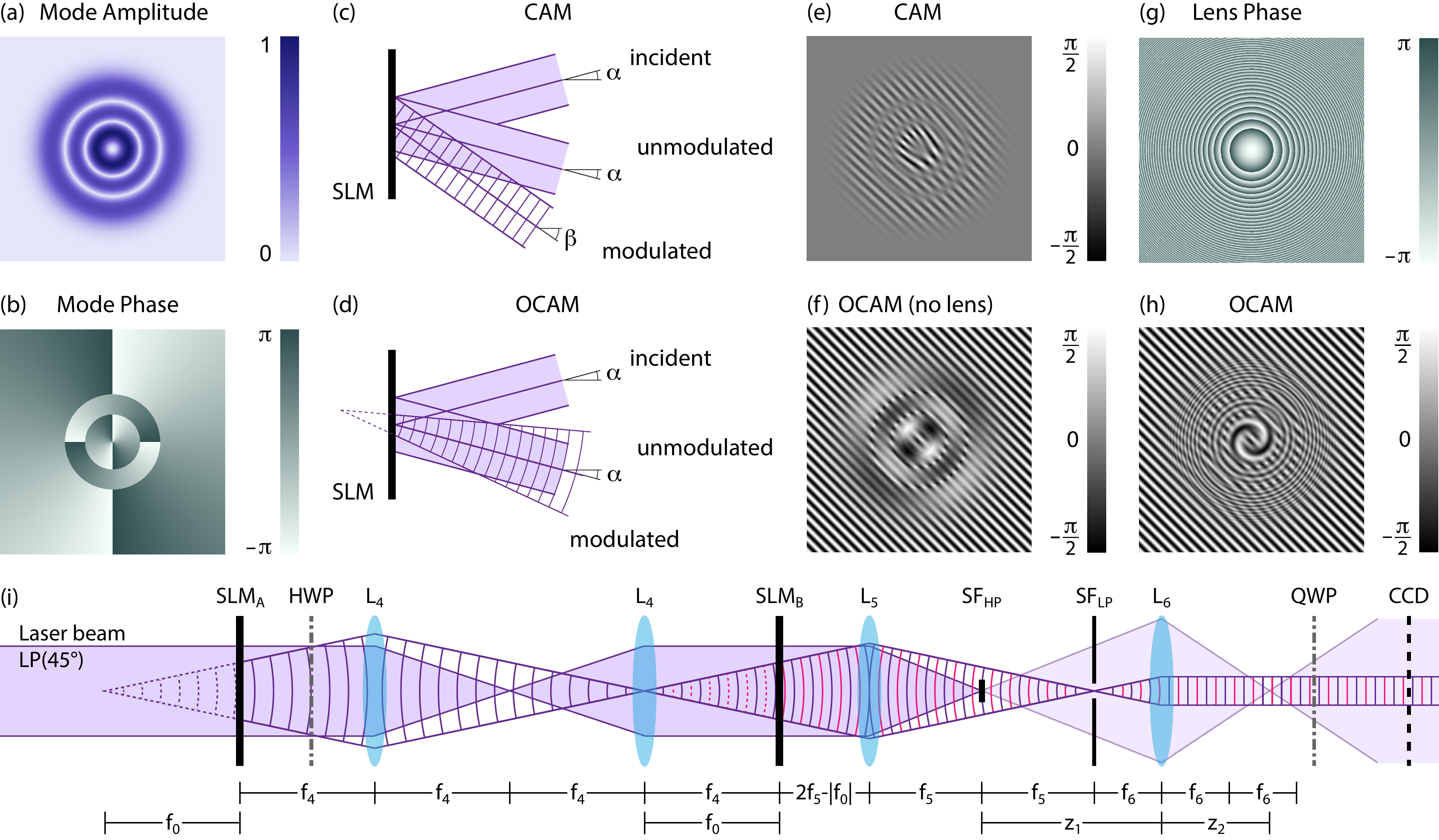}
    \caption{(a) Amplitude and (b) phase of an example beam of interest to be generated. (c) Complex Amplitude Modulation (CAM) basic setup: an incident plane wave is modulated in phase and amplitude by an appropriate hologram to produce an arbitrary optical field at the first diffraction order. (d) On-axis Complex Amplitude Modulation (OCAM) basic setup: an incident plane wave is modulated in phase and amplitude by an appropriate hologram to produce an arbitrary optical field that propagates co-axially to the zeroth order. Corresponding holograms for the (e) CAM and (f) OCAM methods. (g) When using OCAM, the separation of the modulated beam from the zeroth order is achieved by adding to the field of interest a digital divergent lens, resulting in (h) an OCAM hologram that commonly features a spiral pattern. (i) Proposed experimental arrangement to generate vector beams using OCAM. The horizontal component of an initially linearly polarized (at $45^\circ$) plane wave is modulated by a first SLM (SLM$_\text{A}$) to produce a divergent beam (purple lines) while the vertical component remains unchanged (solid area). A Half Wave-Plate (HWP) rotates the polarization by $90^\circ$, and a telescope (lenses L$_4$ in a 4f configuration) is used to image the first SLM plane onto a second one (SLM$_\text{B}$), such that the initially vertically polarized is now modulated (pink lines) and diverges out in the same manner as the initially horizontally polarized component. The focal lengths of the L$_4$ lenses ($f_4$) and the digital one ($f_0$) must be the same, to ensure the two polarization components co-propagate after SLM$_\text{B}$. The un-modulated light (solid area) is filtered out by means of two pupils: one located at the back focal plane of lens L$_5$ (SP$_\text{HP}$) that blocks the central part of the beam and a second one that cleans the beam of interest that is located at $2f_5$ (SP$_\text{LP}$). The vector beam is then re-collimated by lens L$_6$, its polarization optionally changed to the circular basis by using a Quarter Wave-Plate (QWP), and finally it is imaged by a digital camera (CCD). Any remaining un-modulated light would focus at a distance $z_2$ from L$_6$, thus this plane is better avoided.}
    \label{fig:principle}
\end{figure*}

In this section, we describe the working principle behind On-axis Complex Amplitude Modulation (OCAM). Our approach involves two parts: the creation of the OCAM hologram and the separation of the modulated light from the zeroth order, which we will explain in detail in the following.

CAM enables the generation of arbitrary scalar complex ﬁelds with high quality. It uses Computer-Generated-Holograms (CGHs) to encode both the amplitude and phase of the optical fields and while it was initially developed for phase-only spatial light modulators \cite{Davis1999,Arrizon2003}, it has been extended to amplitude-only spatial light modulators \cite{Mirhosseini2013}. There are several schemes to implement CAM, being the ones developed by Arriz\'on \textit{et al.} \cite{arrizon2007} and Bolduc \textit{et al.} \cite{Bolduc2013} two of the most used ones. While our technique can be implemented with either of them, we choose the former. In this case, given a complex field of the form
\begin{equation}\label{Eq:U0}
    U(\vect{r}) = A(\vect{r}) \exp[i\varphi(\vect{r})],
\end{equation}
the Arriz\'on-Type 3 phase modulation associated with the CGH is given by\cite{arrizon2007}
\begin{equation}\label{Eq:cam}
    \psi_\text{CAM}[A(\vect{r}),\varphi(\vect{r})] = \text{J}_1^{\langle-1\rangle}[a\:A(\vect{r})] \,\sin[\varphi(\vect{r})+\text{G}(\vect{r})]
\end{equation}
where $\vect{r}$ represents the transverse coordinates, $(\cdot)^{\langle-1\rangle}$ represents numerical inversion, $\text{J}_1[\cdot]$ is the Bessel function of the first kind of order $1$, and $\text{G}(\vect{r})=2\pi( ux + vy )$ is a blazed grating with spatial frequencies $(u,v)$. Since $\psi_\text{CAM}$ must be a single-valued function, $aA(\vect{r})$ can take only values in the range from zero to $\text{J}_1$'s first maximum ($\text{J}_1[x_1=1.84=0.5857
\pi]=0.5819$), \textit{i.e.} $aA\in[0,0.5819]$, which assuming the field's amplitude is normalized requires $a = 0.5819$, resulting in $\psi_\text{CAM}\in[-x_1,x_1]$. The grating in Eq.~\ref{Eq:cam} is crucial since it diffracts the modulated light away from the zeroth order so that the latter can be removed by means of a spatial filter.

As a way of example, Fig.~\ref{fig:principle} a,b represent respectively the amplitude and phase of a scalar optical mode to be generated. Specifically, a Laguerre-Gaussian beam with $\ell = 2$ and $p=2$ \cite{Rosales2017}. Following the CAM technique (Fig.~\ref{fig:principle}c), a plane wave illuminates an SLM displaying a digital hologram, resulting in a modulated beam propagating at an angle $\beta$ from the horizontal (as depicted in the schematic drawing, although in general it can be deflected in two dimensions) and an un-modulated beam (specular reflection), shown as striped and solid areas, respectively. The appropriate CAM hologram to generate this field is calculated using Eq.~\ref{Eq:cam} and shown in Fig.~\ref{fig:principle}e, where the grating and amplitude modulation are clearly identifiable.

While the grating on the CAM holograms allows generating scalar beams of remarkable quality, it prevents an on-axis implementation of vector beams. These are generated by the superposition of two fields with orthogonal polarizations, thus two modulations, one for each polarization, are necessary to implement them. This has forced experimentalists to either use phase-only holograms in an on-axis configuration, at the cost of reduced beam quality, or CAM together with interferometric approaches, at the cost of stability. In order to overcome this, On-axis Complex Amplitude Modulation (OCAM) changes the purpose of the grating from directing light into the desired diffraction order to reject unwanted energy into said order. In addition, and since it is always necessary to separate the modulated light from the zeroth order, we add a quadratic phase (a digital lens) to the phase of the optical field of interest \cite{christmas2007,Zhang2009}. As a result, the modulated beam out of the SLM diverges and propagates along the same optical axis as the zeroth order (Fig.~\ref{fig:principle}d).
Mathematically, the modulation phase associated with the OCAM hologram is therefore
\begin{equation}\label{Eq:ocam}
    \eqalign{\psi_\text{OCAM}[A(\vect{r}),&\varphi(\vect{r})] = \frac{\text{J}_1^{\langle-1\rangle}[aA(\vect{r})] \sin[\varphi(\vect{r})+\varphi_\text{lens}(\vect{r})]}{2} \\
                        &  + \frac{\text{J}_1^{\langle-1\rangle}[a(1-A(\vect{r}))] \sin[\text{G(\vect{r})}]}{2}  - \frac{x_1}{2} \,,}
\end{equation}
where the phase profile of a lens is $\varphi_\text{lens}(\vect{r}) = -\frac{i\pi}{\lambda f_0}|\vect{r}|^2$, with $f_0$ the focal length and $\lambda$ the wavelength.  The first term in Eq.~\ref{Eq:ocam} corresponds to the on-axis modulation of the beam of interest (Eq.~\ref{Eq:U0}) together with the digital lens, the second term rejects light in a complementary fashion to the desired amplitude, and the extra constants ensure $\psi_\text{OCAM}$ remains in the interval $[-x_1,x_1]$ as necessary. As a reference, Fig.~\ref{fig:principle}f shows the corresponding OCAM hologram with no lens. Here, the grating is visible in regions where the amplitude of the mode is close to zero. When the digital lens (Fig.~\ref{fig:principle}g) is incorporated, the hologram features a spiral pattern (Fig.~\ref{fig:principle}h).

\begin{figure*}
    \centering
    \includegraphics[width=.95\textwidth]{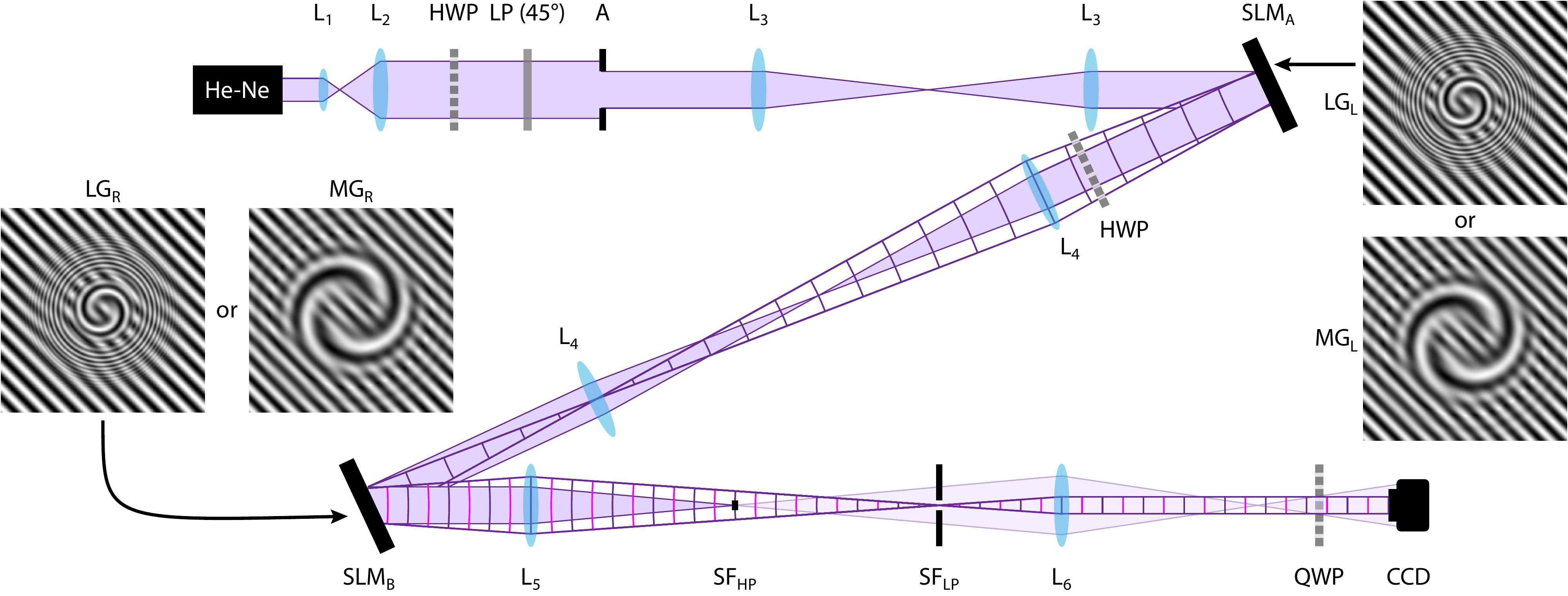}
    \caption{Schematic representation of the experimental setup implemented to demonstrate the proposed technique. An expanded and collimated laser beam ($\lambda=\SI[]{632}{\nano\meter}$) with diagonal polarisation impinges onto a Spatial Light Modulator (SLM$_\text{A}$), where only the horizontal component of polarisation is modulated. The state of polarisation of the emerging beam is then rotated by 90 degrees and sent to (SLM$_\text{B}$), where now the other polarisation component is modulated. The resulting beam is spatially filtered to remove undesired un-modulated light by means of high- (SF$_{\text HP}$) and low-pass (SF$_{\text LP}$) filters. Here and for the sake of clarity, SLM $_{\text A}$ and SLM$_{\text B}$ are represented as two independent SLMs but in our experiment, we used two halves of the same one. Here, L$_{1-6}$ are lenses; HWP and QWP represent a Half and a Quarter Wave-Plates, respectively; LP is a linear polarizer.}
    \label{fig:experiment}
\end{figure*}

Fig.~\ref{fig:principle}i illustrates the proposed implementation of vector modes using OCAM with liquid-crystal SLMs. A collimated laser beam with diagonal linear polarization impinges on a first SLM (A), where only its horizontal component is modulated (purple striped area) while the vertical one is unaffected (solid area). A telescope in a 4f configuration images this plane onto a second SLM (B) and a half wave-plate (HWP) is placed to rotate the polarization by $90^\circ$, such that the initially horizontally polarized light becomes vertically polarized, and vice-versa. Thus the former is simply transmitted by the second SLM, and the latter is modulated by the displayed hologram on SLM$_B$ (pink striped area). It is important for the focal lengths of the real lenses in the telescope ($f_4$) and the digital ones ($f_0$) to be equal, so that the orthogonally polarized modulated beams co-propagate after the SLM$_B$. A mathematical description using the Jones formalism of our method can be found in the supplementary material. Finally, in order to filter out the zeroth order (solid area), two spatial pupils are used. The first one (SP$_\text{HP}$) blocks the center at the back focal plane of SLM$_B$ and should remove as much as possible of the zeroth order while affecting as little as possible the beam of interest. Its size is then critical and must be chosen carefully. The second one (SP$_\text{LP}$) is placed at a plane $2f_5$ away from L$_5$ where the vector beam focuses, since it is arranged in a 2f configuration. Finally, an additional lens (L$_6$) is used to recollimate the vector beam and a quarter wave-plate (QWP) to convert from the linear polarization basis to the circular one, after which the beam can be imaged using a digital camera (CCD). 
Although the zeroth order after the two pupils should be negligible, it is important to consider where any remaining light would focus. Following a simple geometrical optics model, the lens equation establishes that
\begin{equation}\label{Eq:z2}
z_2 = \frac{z_1f_6}{z_1+f_6} = \left( 1+\frac{f_6}{f_5} \right)f_6\,,
\end{equation}
where $|z_1| = f_5+f_6$. Eq.~\ref{Eq:z2} determines the transversal plane to the optical axis where the zeroth order will be the most visible, after which it will diverge. The camera should then be placed after this point. While for the sake of clarity the SLMs in the schematic are transmissive ones, the same idea applies to reflective SLMs, which are usually preferred due to their higher efficiency. Then, by combining the OCAM hologram with the experimental arrangement shown here it is possible to generate vector beams on-axis with high quality and stability. 

\section{Experimental details}
To demonstrate our technique experimentally, we implemented the experimental setup schematically illustrated in Fig.~\ref{fig:experiment}, which is described next. To begin with, a collimated and expanded He-Ne laser beam at a wavelength $\lambda=\SI[]{632}{\nano\meter}$, to form a flat-wave front, impinging onto a liquid crystal SLM (Holoeye Pluto 2.1 Phase Only LCOS with a pixel resolution of $1920\times1080$ pixels and size of \SI[]{8}{\micro\meter}). The beam is expanded through a 10x microscope objective (L$_1$) and the lens L$_2$ with a focal length $f_2=\SI[]{125}{\milli\meter}$. 
The size of the beam impinging on the SLM is controlled by imaging an aperture (A) onto the SLM using the lenses L$_3$ ($f_3=\SI[]{200}{\milli\meter}$) in a 4f configuration, such that the aperture is placed at the front focal plane of the first L$_3$. The polarisation state of the beam impinging on the first SLM (SLM$_\text{A}$) is set to diagonal, with the help of a linear polarizer (LP) and the beam power controlled by rotating a half-wave plate (HWP) located before the polarizer. In this way, only the horizontal polarisation component of the beam is transformed into the desired structured beam, leaving the vertical component unchanged. The resulting beam is imaged onto the second SLM (SLM$_\text{B}$) using lenses $L_4$, both with a focal length $f_4=\SI[]{300}{\milli\meter}$. Here, a second HWP rotates the polarisation state of the beam to transform the horizontal polarisation into vertical and the vertical into horizontal, in this way, now the previously un-modulated light acquires the desired shape, leaving the other component intact. The resulting beam is already a vector beam mixed with the residual un-modulated light that is to be removed with the help of additional lenses, including a digital one. First, the OCAM hologram is multiplexed with a digital diffractive lens of focal length $f_0=\SI[]{-300}{\milli\meter}$ (see \cite{SPIEbook} for details), which causes the modulated light to diverge, leaving the residual (zeroth order) unchanged. A physical lens $L_5$ of focal length $f_5=\SI[]{250}{\milli\meter}$ placed at a distance $2f_0-|f_0|=\SI[]{200}{\milli\meter}$ from the SLM causes the residual un-modulated light to focus at a distance $d_1=f_5$, where a stop of specific diameter is located to block it, and the modulated light to focus at a distance $d_2=2f_5$, where a spatial filter is placed to block higher diffraction orders and clean the generated vector beam. The last element is a lens $L_6$ of focal length $f_6=\SI[]{150}{\milli\meter}$ that collimates the beam to produce the desired vector beam in the linear polarisation basis, and can be transformed to the circular basis by means of a quarter-wave plate.

\begin{figure*}
    \centering
    \includegraphics[width=0.95\textwidth]{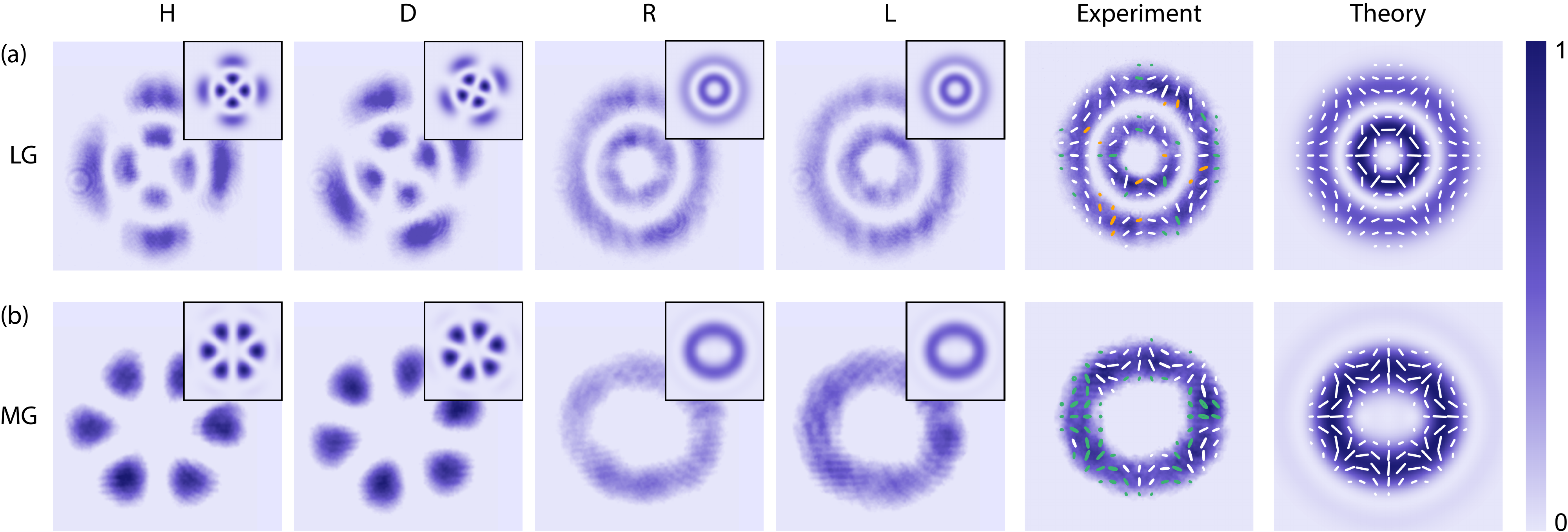}
    \caption{Experimentally generated (a) Laguerre- and (b) Mathieu-Gauss vector modes. The first four panels from left to right show the polarisation components required to reconstruct the transverse polarisation components of the vector modes shown on the last two panels marked as Experiment and Theory. Here, the top-left insets show the numerically-simulated intensities.}
    \label{fig:simulation}
\end{figure*}

We investigated the performance of our approach by creating two examples of vector modes from the families Laguerre- and Helical Mathieu-Gaussian modes, both embedded with Orbital Angular Momentum (OAM). Corresponding holograms are shown as insets in Fig.~\ref{fig:experiment}, where the typical grating and spiral patterns are observed. Notice how in this case the holograms in both SLMs are the same since the sign of the OAM from the first modulation is inverted by the reflection on the second SLM. This is, however, not the general case since the two holograms are completely independent so that any vector beam can be produced. A qualitative analysis of the generated beams was performed by reconstructing their transverse polarisation distribution through Stokes polarimetry, using a series of polarisation filters and a Charged-Coupled Device (CCD) camera from Thorlabs, with a spatial resolution of $1280\times1024$ pixels and pixel size \SI[]{5.3}{\micro\meter} (see \cite{Rosales2021} for details about Stokes polarimetry).

\section{Results} 
To validate the performance of our technique, in this section, we show a representative selection of experimental results, including examples from two different sets of solutions to the wave equation, namely, the Laguerre- and Mathieu-Gauss vector beams. As a first test, we analyze the quality of the generated vector modes by means of Stokes polarimetry, whereby a comparison of the transverse polarisation distribution of our experimental results and theory was performed. As a second test, we analyze the stability of the vector modes as a function of time. 

To begin with, we generated vector modes from the well-known set of Laguerre-Gaussian (LG) modes. Mathematically, they are  described by the relation \cite{Rosales2018Review},
\begin{equation}\label{PSPol2}
\eqalign{
{\bf LG}_{p_1,p_2}^{\ell_1,\ell_2}(\mathbf{r})=\cos\phi\,&\text{LG}_{p_1}^{\ell_1}(\mathbf{r})\mathbf{\hat{e}_R}\\ &+\sin\phi\, \text{LG}_{p_2}^{\ell_2}(\mathbf{r})\e^{i\theta}\mathbf{\hat{e}_L}\,,}
\end{equation}
where $\text{LG}_{p}^{\ell}$ are the Laguerre-Gaussian modes with radial and azimuthal indices $p$ and $\ell$, respectively, with weighting coefficients specified by the parameter $\phi$. The unitary vectors $\mathbf{\hat{e}_R}$ and $\mathbf{\hat{e}_L}$ represent the right- and left-handed circular polarisations, respectively. Finally, the parameter $\theta$ is the intermodal phase that rotates the polarisation structure of vector modes and is the main responsible for the polarisation fluctuations in vector beams generated by interferometric means. Fig.~\ref{fig:simulation}a shows an experimentally generated LG vector mode with parameters $p_1=p_2=1$, $\ell_2=-\ell_1=2$, $\phi=\frac{\pi}{4}$ and $\theta=0$. The first four panels, from left to right, show the intensities required to reconstruct the transverse polarisation of the beam, and the top-right insets show corresponding theoretical intensities. The last two panels show the experimentally and theoretically reconstructed transverse polarisation distribution, overlapped onto the intensity profile. Here, white lines represent linear polarisation, and orange and green ellipses represent right- and left-handed elliptical polarisation, respectively.

As a second example, we generated helical Mathieu-Gauss vector modes, which are constructed as a non-separable superposition of circular polarisations and helical Mathieu-Gauss beams as \cite{rosales2021Mathieu},
\begin{equation}\label{HMVB}
    \eqalign{
    {\bf MG}_{m_1,m_2}(\mathbf{r}) = \cos\phi\,&\text{MG}^+_{m_1}(\mathbf{r}) \mathbf{\hat{e}_R}\\ &+ \sin\phi\,\text{MG}^-_{m_2}(\mathbf{r})\e^{i\theta}\mathbf{\hat{e}_L},}
\end{equation}
where the functions $\text{MG}^+_{m_1}$ and $\text{MG}^-_{m_2}$ represent the finite-energy helical Mathieu-Gauss (hMG) beams of order $m \in [0, 1, 2, 3, ...]$ and defined eccentricity $e \in [0,1]$, a parameter that controls the ellipticity of the mode. Incidentally, such beams are non-diffracting over the finite propagation interval $[-z_{max},z_{max}]$, where $z_{max}=\omega_0k/k_t$, with $\omega_0$ the beam waist of the Gaussian beam, $k=2\pi/\lambda$ the wave number and $k_t$ the component of the wave vector along the transverse plane. In a similar way to the previous case, Fig. \ref{fig:simulation}b shows the experimental results obtained for the hMG vector beam with parameters $m_1=m_2=3$, opposite helicity, $e=0.3$ and $k_t=5$, compared to the theoretical mode. Notice that in both cases shown in Fig.~\ref{fig:simulation}, the beams show high quality in both their intensity profile and transverse polarisation distribution. Small deviations of the experimental results from the theoretical predictions are prominently caused by the fact that SLM screens are not completely flat.

\begin{figure}
    \centering
   \includegraphics[width=0.47\textwidth]{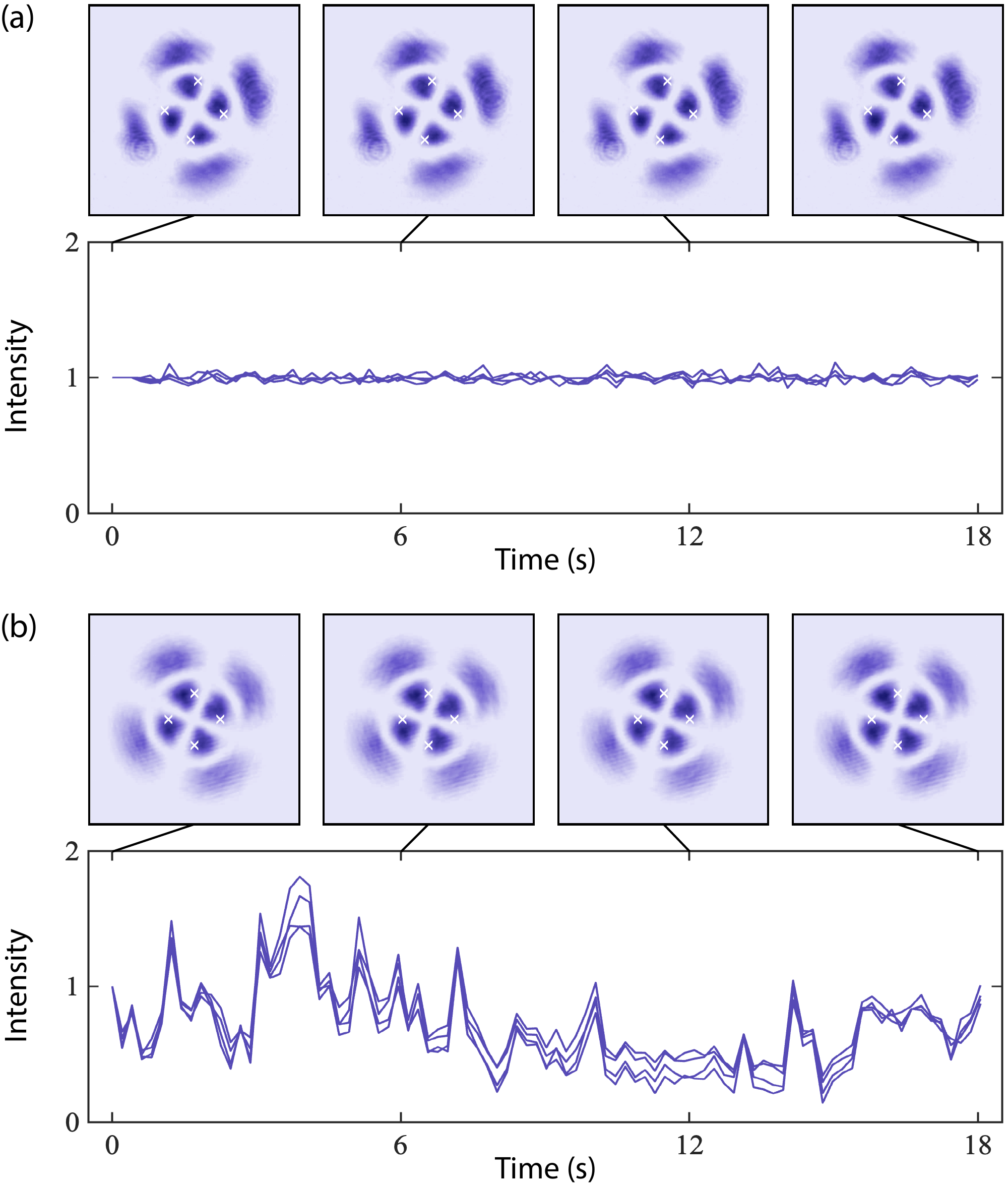}
    \caption{Comparative analysis of the stability of (a) OCAM (Multimedia file 1) with (b) an interferometric generation  (Multimedia file 2). The intensity variations of OCAM are very low compared to the other technique.}
    \label{fig:stability}
\end{figure}

To analyze the stability of the vector beams produced with our technique, we recorded videos of the generated vector beam passing through a linear polarizer (See Multimedia files 1 and 2). 
Specifically, we compared our technique against another generation technique, employing a common-path Sagnac interferometer, one of the most stable interferometric methods \cite{Perez-Garcia2017,Perez-Garcia2022}. At this stage, it is worth reminding that the non-stability arises from the fact that both beams travel along different paths and therefore, the perturbations are not common to both beams causing the intermodal phase to change randomly. As a result, the transverse polarisation distribution of the vector beam also oscillates randomly. Mathematically, consider a vector mode of the form defined in Eq.~\ref{PSPol2}.  At a given point $(\rho_0, \phi_0)$, the beam is given by
\begin{equation}
    \eqalign{
    {\bf LG}_p^{\ell}(\rho_0, \phi_0) =\, U_0(\rho_0)&e^{i\ell\phi_0}\hatt{e}_R\\ &+ e^{i\delta}U_0(\rho_0)e^{-i\ell\phi_0}\hatt{e}_L\,,}
\end{equation}
where $\delta\sim \text{r.v.}$ is an intermodal phase arising from random fluctuations in the path of both beams, caused by, for instance,  mechanical vibrations of the optical elements (mirrors, beam splitters, etc.).  The effect of this term can be observed by projecting the vector beam along a given polarisation direction, which produces a petal-like intensity distribution (such as the ones shown in the first two panels of Fig.~\ref{fig:simulation}), expressed as $\vect{E}(\rho_0, \phi_0) = \mathbf{J_{LP}}\,{\bf LG}_p^{\ell}(\rho_0, \phi_0)$, where $\mathbf{J_{LP}} = \frac{1}{2}\left[{\begin{array}{cc}
    1 & 1\\
    1 & 1
  \end{array} }\right]$ is the Jones matrix (in circular polarization basis) of a linear polarizer oriented horizontally.  The intensity at the point $(\rho_0, \phi_0)$ is therefore
\begin{equation}\label{Eq:int}
	I(\rho_0, \phi_0) = 4|U_0(\rho_0)|^2\cos^2(\phi_0 + \delta)\,.
\end{equation}
From Eq.~\ref{Eq:int} is clear that the random phase term $\delta$ causes the petal--like intensity structure to fluctuate. In the case of a non-stable vector beam, such petal-like structure oscillates randomly over time. Crucially, in our technique these oscillations are negligible. To demonstrate this, we tracked the intensity at four different positions in the transverse plane of the beam and over a time-lapse of 18 seconds. Fig.~\ref{fig:stability} shows the intensity variations of the petal structure for both cases mentioned above, comparing the performance of our technique (Fig.~\ref{fig:stability}a) against the interferometric one (Fig. \ref{fig:stability}b). Notice the high variations of the latter, compared to the first, showing that indeed our technique is highly stable. To quantify the stability in both approaches, we computed the standard deviation at the aforementioned fixed positions, over the time interval composed of 100 frames.  The measurement results yield $\sigma_\text{Sagnac} = 0.34$ for the interferometric setup, and $\sigma_\text{OCAM} = 0.03$ for our scheme, thus confirming the visual perception. 

\section{Discussion and conclusions} 
Complex vector modes have become the subject of study in various research fields where their unique properties are pioneering novel applications. Examples include optical tweezers, as it has been demonstrated theoretically and experimentally that radially polarised vector beams possess a stronger longitudinal optical force, compared to linearly-polarized scalar beams \cite{Michihata2009,Bhebhe2018}. In optical metrology, the nonseparability of vector beams has been used as a real-time tracking tool for two- and three-dimensional sensing of moving objects \cite{BergJohansen2015,hu2019}. In classical and quantum communications, the orthogonality property of vector beams provides alternative means to encode information in order to increase the security of information transfer or to perform real-time correction in quantum communications \cite{Ndagano2018,Ndagano2017,milione2015using}. More recently, an increasing interest in the propagation dynamics of vector beams is gaining popularity. Examples include the case of vector modes that upon propagation oscillate from one vector state to another or evolve into quasi-scalar beams, either along the transverse or the longitudinal direction \cite{Otte2018,Davis2016,zhong2021,PengLi2018,Hu_2021,Hu2023}. In all the above examples, it is evident the need for generation techniques capable of generating arbitrary vector beams with high quality, high stability, and in a flexible manner. To date, these requirements are only partially fulfilled. On the one hand, q-plates and J-plates offer high stability but at the expense of no flexibility, allowing the generation of a unique vector beam \cite{Marrucci2006}. On the other hand, techniques based on spatial light modulators, where vector beams are commonly generated through interferometric arrays, offer high quality and flexibility but at the expense of low stability. As such, in this manuscript, we presented a non-interferometric technique capable of generating high-quality arbitrary vector beams with very high stability. The technique relies on the use of a liquid crystal Spatial Light Modulator (SLM), which allows the generation of arbitrary complex vector beams in phase and amplitude \cite{arrizon2007}. For its implementation, we take full advantage of the fact that SLMs can only modulate a linear polarisation component of light, commonly the horizontal. Further, the screen of the SLM can be digitally split into two independent sections, each of which encodes a specially designed hologram provided with complex amplitude modulation. An expanded and collimated diagonally polarised beam, which for practical purposes can be seen as the superposition of two beams, one with horizontal polarisation and another with vertical, is sent to the first half of the SLM. As a result, only the horizontal polarisation acquires the modulation imposed on the hologram, leaving the vertical intact. Afterward, both polarisation components are rotated so that the vertical component becomes horizontal and the horizontal component becomes vertical. The beam is redirected to the second section of the SLM to modulate now the other polarisation component. As a result of this sequential modulation, we obtain a high-quality vector beam. Similar configurations have been implemented before but our technique incorporates a nontrivial on-axis complex amplitude modulation hologram, which highly improves the quality of the generated beams. Moreover, since both beams travel always along the same optical path, the generated beam is highly stable, as we demonstrated in our experimental results section. We anticipate that our technique will be of great relevance in applications that rely on vector beams with high stability, for example, in optical metrology to determine the physical properties of a given medium after its interaction with a vector beam.

\section*{Data availability statement}
The data that support the findings of this study are available upon request to the authors.

\section*{Disclosures}
The authors declare that there are no conflicts of interest related to this article.  
\section*{References}
\bibliographystyle{iopart-num}
\providecommand{\newblock}{}

\end{document}


\title[On-axis complex-amplitude modulation for super-stable vector modes generation]{On-axis complex-amplitude modulation for super-stable vector modes generation\\Supplementary Material}


\author{Valeria Rodr\'iguez-Fajardo$^1$, Fernanda Arvizu$^2$, Dayver Daza-Salgado$^2$, Benjamin Perez-Garcia$^3$, Carmelo Rosales-Guzm\'an$^2$}
\address{$^1$Department of Physics and Astronomy, Colgate University, 13 Oak Drive, Hamilton, NY 13346, United States of America}
\address{$^2$Centro de Investigaciones en Óptica, A.C., Loma del Bosque 115, Colonia Lomas del campestre, 37150 León, Gto., Mexico}
\address{$^3$Photonics and Mathematical Optics Group, Tecnologico de Monterrey, Monterrey 64849, Mexico.}

\ead{vrodriguezfa@gmail.com}

\vspace{10pt}
\begin{indented}
\item[]\today
\end{indented}

%
%
%

%
%
%
%

\section{Complex modulation of scalar beams} \label{sec:complexmodulation}

Consider the Jones matrix of a phase--only SLM, with its extraordinary axis parallel to the $x$--axis
\begin{equation}
  J_{\text{SLM}}(x,y) =
  \left[ {\begin{array}{cc}
    \exp[i\delta(x,y)] & 0\\
    0 & 1
  \end{array} } \right],
\end{equation}
where $\delta(x,y)$ is the phase corresponding to each pixel located at the coordinate $(x,y)$.  Let us also assume an input plane wave horizontally polarized
\begin{equation}
  \vect{E}_0(x,y) =
  \left[ {\begin{array}{c}
    1\\
    0
  \end{array} } \right].
\end{equation}
The action of the SLM is therefore written as
\begin{equation}
\eqalign{\vect{E}(x,y) &=
    \left[ {\begin{array}{cc}
    \exp[i\delta(x,y)] & 0\\
    0 & 1
  \end{array} } \right] 
  \left[ {\begin{array}{c}
    1\\
    0
  \end{array} } \right],\\
  & = \left[ {\begin{array}{c}
    \exp[i\delta(x,y)]\\
    0
  \end{array} } \right].
  }
\end{equation}
The idea is to encode in $\delta(x,y)$ the information of a complex field
\begin{equation}
	U(x,y) = A(x,y)\exp(i\varphi(x,y)),
\end{equation}
where $A(x,y)$ is the amplitude and $\varphi(x,y)$ is the transverse phase.  A phase carrier is added to the field $U$ such that
\begin{equation}
	V(x,y) = \exp[i(k_{0x} x + k_{oy} y)]\: U(x,y),
\end{equation}
with $(k_{0x}, k_{0y})$ spatial frequencies.  Following Arrizon's type--3 method, we calculate
\begin{equation}
\delta(x,y) = f(|V(x,y)|)\sin(\Arg(V(x,y))),
\end{equation}
where $f(\cdot)$ comes from the inversion of the Bessel function $\text{J}_1(\cdot)$ and $\Arg(\cdot)$ computes the principal argument.  Note that the desired output beam $U(x,y)$ is ''tilted'' according to $(k_{0x}, k_{0y})$.  We can design a suitable filter to separate the desired mode from the unwanted light.  This is written as 
\begin{equation}
	\eqalign{\tilde{\vect{E}}(k_x,k_y) &= \mathcal{F}[\vect{E}(x,y)],\\
	& = \left[ {\begin{array}{c}
    \mathcal{F}[\exp[i(f(|V(x,y)|)\sin(\Arg(V(x,y))))]]\\
    0
  \end{array} } \right]\\
  & = \left[ {\begin{array}{c}
  \tilde{U}(k_x-k_{0x}, k_y-k_{0y}) + \text{unwanted light} \\
  0 \end{array} } \right].
}
\end{equation}
The spatial filtering is achieved by applying a circ$(\cdot)$ function, that is
\begin{equation}
  \eqalign{\tilde{\vect{E}}(k_x,k_y) = \\ \left[ {\begin{array}{c}
  \text{circ}\left(\frac{k_x-k_{0x}}{R}, \frac{k_y-k_{0y}}{R}\right)\left[\tilde{U}(k_x-k_{0x}, k_y-k_{0y}) + \text{unwanted light}\right] \\
  0 \end{array} } \right],\\
  = \left[ {\begin{array}{c}
  \tilde{U}(k_x-k_{0x}, k_y-k_{0y}) \\
  0 \end{array} } \right].
}
\end{equation}
Finally, we take the inverse Fourier transform
\begin{equation}
	\eqalign{\vect{E}_f(x,y) &= \mathcal{F}^{-1}[\tilde{\vect{E}}(k_x,k_y)],\\
	&= \left[ {\begin{array}{c}
   \exp[i(k_{0x} x + k_{oy} y)]\:U(x, y) \\
  0 \end{array} } \right].}
\end{equation}

\section{On--axis generation of vector beams}\label{sec:vectorbeams}

Consider the Jones matrix of a phase--only SLM and the illumination of a diagonally polarized input plane wave
\begin{equation}
\eqalign{\vect{E}(x,y) &=
    \frac{1}{\sqrt{2}}\left[ {\begin{array}{cc}
    \exp[i\delta_1(x,y)] & 0\\
    0 & 1
  \end{array} } \right] 
  \left[ {\begin{array}{c}
    1\\
    1
  \end{array} } \right],\\
  & = \frac{1}{\sqrt{2}}\left[ {\begin{array}{c}
    \exp[i\delta_1(x,y)]\\
    1
  \end{array} } \right].
  }
\end{equation}

Then, a HWP with its fast--axis oriented at 45$^\circ$ transforms the field to
\begin{equation}
\eqalign{\vect{E}_2(x,y) &=
    \frac{1}{\sqrt{2}}\left[ {\begin{array}{cc}
    0 & 1\\
    1 & 0
  \end{array} } \right] 
  \left[ {\begin{array}{c}
    \exp[i\delta_1(x,y)]\\
    1
  \end{array} } \right],\\
  & = \frac{1}{\sqrt{2}}\left[ {\begin{array}{c}
	1\\
    \exp[i\delta_1(x,y)]
  \end{array} } \right].
  }
\end{equation}

After the rotation, we illuminate another SLM
\begin{equation}\label{eq:naive}
\eqalign{\vect{E}_f(x,y) &=
    \frac{1}{\sqrt{2}}\left[ {\begin{array}{cc}
    \exp[i\delta_2(x,y)] & 0\\
    0 & 1
  \end{array} } \right] 
  \left[ {\begin{array}{c}
    1\\
    \exp[i\delta_1(x,y)]
  \end{array} } \right],\\
  & = \frac{1}{\sqrt{2}}\left[ {\begin{array}{c}
    \exp[i\delta_2(x,y)]\\
    \exp[i\delta_1(x,y)]
  \end{array} } \right].
  }
\end{equation}
Please note that $\vect{E}_f(x,y)$ features different polarisation states across the transverse plane, according to the phases $\delta_j(x,y)$ that can be independently shaped.  Also note that the amplitude in both components is the same --corresponding to the initial input beam--, limiting the set of polarization states that can be generated.  Finally, the basis of Eq.\ \ref{eq:naive} is $(\hat{H}, \hat{V})$, but can be easily changed to the circular basis via a QWP after the second SLM.

\section{OCAM}

The basic notion behind this approach is to combine the techniques outlined in Sec.\ \ref{sec:complexmodulation} and \ref{sec:vectorbeams}.  The idea is to perform complex modulation in each component, while remaining on--axis propagation.  

Let us remind Eq.\ \ref{eq:naive} and consider it as an starting point for our OCAM implementation
\begin{equation*}
\eqalign{\vect{E}_f(x,y) = 
    \frac{1}{\sqrt{2}}\left[ {\begin{array}{c}
    \exp[i\delta_2(x,y)]\\
    \exp[i\delta_1(x,y)]
  \end{array} } \right].
  }
\end{equation*}
Here, $\delta_j(x,y)$ is the corresponding phase that generates a field
\begin{equation}
V_j(x,y) = \exp[ik\rho^2/2f]\:U_j(x,y),
\end{equation}
where $U_j(x,y) = A(x,y)\exp[i\varphi_j(x,y)]$ is the desired complex field.

To compute $\delta_j(x,y)$ we used Arrizon's type--3 method in the following way
\begin{equation}
	\eqalign{\delta_j(x,y) = &\frac{1}{2}\text{J}^{-1}(a\:A_j(x,y))\sin[\varphi_j(x,y) + k\rho^2/2f_j] + \\
	&\frac{1}{2}\text{J}^{-1}(a\:(1-A_j(x,y)))\sin[k_{0x}x+k_{0y}y] - \frac{x_1}{2}},
\end{equation}
where the first term generates the desired mode with a quadratic phase dependence, that is, the beam will either converge or diverge, depending on the sign of $f_j$.  The second term involves the portions of light that do not overlap over the desired mode.  These portions are thrown away via the added linear grating to avoid perturbing the beam.  A system of lenses is carefully positioned to match both modulations in the same propagation plane.  Be aware also, that unwanted modulated light indeed exists in both modulations and thus have to be filtered.  To achieve the desired filter, we perform a spatial long--pass Fourier filter, i.e.\ we blocked the on--axis intensity in the focal plane after a lens.